\documentclass[usegraphicx,usenatbib]{mn2e}

\author[A. Kaviani et al.]  {A.~Kaviani$^1$,  M.  G.~Haehnelt$^2$  and
G.~Kauffmann$^3$\\ $^1$Astrophysics Group, Blackett Laboratory,
Imperial College, Prince Consort Road, London SW7 2BW, UK\\
$^2$Institute of Astronomy, University of Cambridge, Madingley Road,
Cambridge CB3 0HA, UK\\ $^3$Max-Planck  Institut f\"{u}r  Astrophysik,
D-85748 Garching, Germany}

\title[SCUBA sources: hot starbursts or cold extended galactic
dust?]{Modelling SCUBA sources in a $\Lambda$CDM cosmology: hot
starbursts or cold extended galactic dust?}

\usepackage{aas_macros}
\usepackage{amssymb}
\begin{document}
\maketitle

\begin{abstract}
Previous modelling has demonstrated that it is difficult to reproduce
the SCUBA source counts within the framework of standard hierarchical
structure formation models if the sources are assumed to be the
high-redshift counterparts of local ultra-luminous infrared galaxies
with dust temperatures in the range 40 -- 60 K. Here, we show that the
counts are more easily reproduced in a model in which the bulk of the
sub-millimetre emission comes from extended, cool (20 -- 25 K) dust in
objects with star formation rates of 50 -- 100
$\textrm{M}_{\odot}\textrm{yr}^{-1}$. The low temperatures imply
typical sizes of $\sim 1 (S_{850}/1\textrm{mJy})^{1/2}$ arcsec, a factor
two to three larger than those predicted using starburst-like spectral
energy distributions. Low dust temperatures also imply a ratio of
optical/UV to 850-$\mu$m flux which is 30 -- 100 times smaller, for
the same optical depth, than expected for objects with a hot,
starburst-like SED. This may help explain the small overlap between
SCUBA sources and Lyman--break galaxies. 
\end{abstract}


\section{Introduction}
Studies of galaxy properties in the infrared (IR) have added much to
our understanding of how and when galaxies form their stars
(e.g. \citealt{1997MNRAS.289..490R}; \citealt{1997AJ....114...54M};
\citealt{1998hdf..symp..219D}; \citealt{1998ApJ...505L.103L};
\citealt{1998ApJ...508..539P}; \citealt{1999ApJ...517..148F};
\citealt*{1999ApJ...521...64M}; \citealt{1999ApJ...519....1S}). At low
redshifts, the bright end of the infrared galaxy luminosity function
is dominated by strongly starbursting systems, which show a rapid
increase in space density towards larger redshifts (see review by
\citealt{1996ARA&A..34..749S}). The large number of bright
submillimetre (sub-mm) sources that were detected by the SCUBA
instrument on the JCMT nevertheless came as a surprise
(\citealt*{1997ApJ...490L...5S}; \citealt{1998Natur.394..241H};
\citealt*{1999ApJ...518L...5B}; \citealt{1999ApJ...512L..87B};
\citealt{2000AJ....120.2244E}; \citealt{2002MNRAS.331..817S}). If the
the sub-mm sources are the high-redshift equivalent of the strong
starbursts observed at low redshifts, the high number counts imply
that strong evolution in the space density of bright IR sources would
have to continue beyond $z\sim$ 3 -- 4.

Many authors have devised phenomenological models to reproduce the
sub-mm source counts. These models assume a local luminosity
function of the sources and a parametrized law for the evolution of the
sources in luminosity and/or density that is constrained to fit the
counts at various wavelengths (see \citealt{2001ApJ...549..745R} and
references therein).

It is nevertheless important to understand the observed sub-mm counts
within the standard cosmological paradigm. Previous modelling based on
semi-analytic (\citealt{1998MNRAS.295..877G};
\citealt{2000A&A...363..851D}) or gas
dynamical \citep{astro-ph/0107290} simulations, has encountered
considerable difficulties in reproducing the brightest sub-mm
sources. If the bright SCUBA sources are the high-redshift
counterparts of the local ultra-luminous infrared galaxies (ULIRGs),
their dust temperature should be high (40 -- 60 K). They are then
inferred to be forming stars at rates of a few hundred to a thousand
solar masses per year, the exact value depending on the detailed
assumptions regarding extinction, initial mass function and the
duration of the burst. Such extreme star formation rates are very
difficult to achieve in standard hierarchical galaxy formation
models. \citet{2000A&A...363..851D} show that under these assumptions,
they are only able to reproduce the sub-mm counts with a model in
which essentially {\em all} baryons present in massive dark matter
haloes at redshifts $\sim 3$ turn into stars over timescales of order
$10^8$ years. They propose that high merging rates in the early
Universe may be responsible for this extremely efficient conversion of
gas into stars, but do not attempt to model this in detail. On the
other hand, \citet{1999MNRAS.309..715B} have modelled the merging
rates of dark matter haloes in hierarchical cosmologies and find that,
in order to fit the counts, the halo mass-to-infrared light ratio of a
typical galaxy merger must be 200 times smaller at redshift 3 than at
the present day. This is clearly an extreme requirement.

The form of the spectral energy distribution (SED) of the sub-mm
galaxies is, however, a source of major uncertainty in the
predictions. Recently \citet{astro-ph/0107290} modelled the sub-mm
counts using N-body plus smoothed particle hydrodynamics simulations
and showed that better agreement with the data was obtained if the
dust temperatures were at the lower end of the range usually assumed
for starbursting galaxies. This follows suggestions by
\citet{2001ApJ...549..745R} and \citet{cirrus2.ps} that the sub-mm
emission from high-redshift galaxies could arise from cirrus-like
emission of cold extended dust.  An extended distribution of cold dust
may, for example, be caused by wind-driven dust outflows
(\citealt*{1990ApJS...74..833H}; \citealt*{1999A&A...343...51A}).

In this paper, we explore the degeneracy between the inferred star
formation rate and dust temperature explicitly and treat temperature
as a free parameter when fitting the sub-mm counts. We demonstrate
that hierarchical galaxy formation models can reproduce the sub-mm
counts with moderate star formation rates if the sub-mm emission
arises from cool, extended dust and discuss the implications for the
size of the emitting region. We use the galaxy formation model
described in \citet*{1993MNRAS.264..201K} and
\citet{1999MNRAS.303..188K} and the diffuse cirrus emission SEDs of
ERR.

In \S\ref{semianalytic}, we describe the galaxy formation model used
to predict the star formation rates and outline the difficulties of
hierarchical models in accommodating a high space density of strong
starbursts out to large redshifts. \S\ref{SED} discusses our modelling
of the SED and explores the degeneracies in the dust models. In
\S\ref{fit}, we present our results and a comparison with observations
and in \S\ref{discuss} we discuss their implications for the nature of
the sub-mm sources. \S\ref{conculsions} presents our conclusions.

\section{The galaxy formation model}\label{semianalytic}

\subsection{Star formation in hierarchical galaxy formation models} 

We have used the star formation rates predicted by the semi-analytic
galaxy formation model of \citet{1993MNRAS.264..201K}. Here, we briefly
recapitulate some of the assumptions of the model relevant for our
predictions of the sub-mm counts. The model and semi-analytic
prescriptions are described in detail in the above
reference and elsewhere (\citealt{1993MNRAS.262..627L};
\citealt{1999MNRAS.303..188K}; \citealt{1999MNRAS.310.1087S};
\citealt{2000MNRAS.319..168C}). The calculations are for a flat
$\Lambda$CDM model with $\Omega_{\Lambda}=0.7$ and $H_{0}=70$
kms$^{-1}$Mpc$^{-1}$. The model follows the formation and evolution of
galaxies within a merging hierarchy of dark matter haloes. Simple
prescriptions take into account the dissipative cooling of the gas,
the formation of stars, feedback from supernovae and the merging rates
of galaxies located within the same dark matter halo. Star formation
is modelled by the following equation:
\begin{equation}
\dot{M}=\alpha\frac{M_{\rm cold}}{t_{\rm dyn}},
\end{equation}
where $\dot{M}$ is the star formation rate, $M_{\rm cold}$ is the mass
of the available cold gas, $t_{\rm dyn}$ is the dynamical time of the
galaxy, and $\alpha$ is a parameter controlling the efficiency of star
formation. $M_{\rm cold}$ depends on the cooling rate of the hot gas
in the dark matter halo, the star formation rate and the efficiency of
supernovae in heating the gas. In the model, $\alpha$ is a free
parameter that is set to reproduce the luminosity of a Milky Way-type
galaxy. The star formation efficiency is enhanced by a factor  $10$ if
a merging event takes place.  The duration of the period of high
efficiency star formation, $t_b$, scales with the dynamical time of
the galaxy and is another parameter of the model. (See also
\citealt*{astro-ph/0210030})

\subsection{The space density of starbursting galaxies at high
redshift}\label{hierarchy}

\begin{figure}
\begin{center}
\includegraphics[width=1.0\columnwidth]{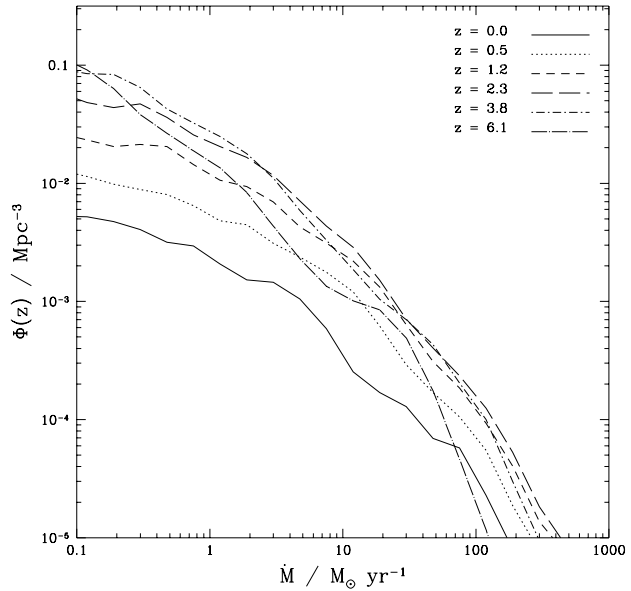}
\end{center}
\caption{The combined star formation rate function at various
redshifts.}
\label{nsfr}
\end{figure}

Fig. \ref{nsfr} shows the space density of galaxies as a function of
star formation rate at a series of different redshifts. Overall, the
typical star formation rates increase towards higher redshift. This is
because the dynamical times of high-redshift galaxies are shorter,
implying higher star formation rates, and because galaxies are
typically more gas-rich at high $z$. However, the space density of
galaxies with star formation rates above 50 -- 100
$\textrm{M}_{\odot}\textrm{yr}^{-1}$ drops at redshifts greater than
2--3. Such systems only occur in the most massive dark matter haloes,
which are rare at early epochs. Objects with star formation rates of a
few hundred to a thousand $\textrm{M}_{\odot}\textrm{yr}^{-1}$ are
very rare at all epochs.

\begin{figure}
\begin{center}
\includegraphics[width=1.0\columnwidth]{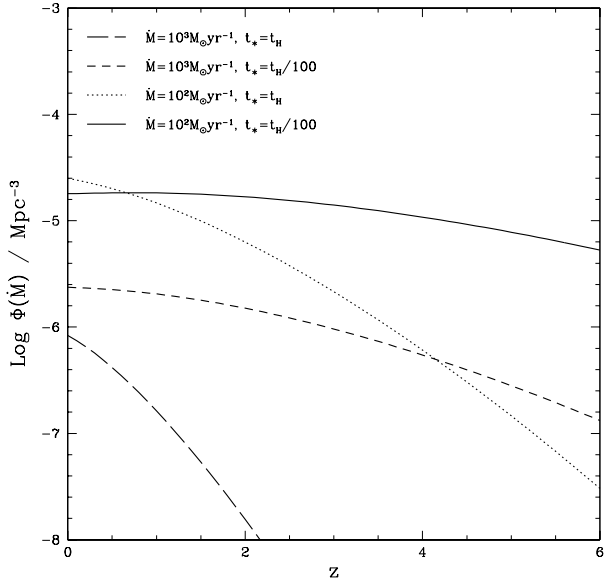}
\end{center}
\caption{The co-moving density of galaxies that produce stars at rates
$10^{2}$ and $10^{3}$ $\textrm{M}_{\odot}\textrm{yr}^{-1}$ as obtained
directly from the Press--Schechter mass function. $t_{\ast}$ is the
star formation timescale, which we have related to the Hubble time,
$t_{\rm H}$, as shown in the figure.}
\label{phivsz}
\end{figure}

To demonstrate that this is a genuine feature of all hierarchical
galaxy formation models and not an artefact of the particular star
formation and feedback prescriptions we have chosen, we have carried
out the following simple exercise. We assign a fiducial star formation
rate to each dark matter halo in the Universe by assuming that at any
given epoch, one tenth of the available baryons in the halo are
forming stars over a timescale $t_{\ast}$. The star formation rate per
halo may then be written as
$\dot{M}=(\Omega_{\ast}/\Omega_{0})(M_{h}/t_{\ast})$, where
$\Omega_{\ast}\sim \Omega_{B}/10$, $\Omega_{B}=0.02h^{-2}$
\citep*{2002Natur.415...54B} and $\Omega_{0}=0.3$ ($\Omega_B$ and
$\Omega_0$ are the baryonic and total co-moving mass densities in the
Universe and $M_{h}$ is the mass of the halo harbouring the
galaxy). We consider two cases: quiescent star formation with
$t_{\ast}= t_{\rm H}$, and bursty star formation with $t_{\ast}=
t_{\rm H}/100$ and calculate halo abundances using the standard
Press-Schechter formula. Fig. \ref{phivsz} shows the redshift
evolution of the space density of objects with different star
formation rates. As can be seen, the space density of massive haloes
with $\sim 10^{3}$ $\textrm{M}_{\odot}\textrm{yr}^{-1}$ is small and
drops very rapidly towards larger redshifts.

\citet{2002MNRAS.331..817S} derived a density $\Phi(\dot{M}\ge 1000$
$\textrm{M}_{\odot}\textrm{yr}^{-1}, z\ge 1.5) \simeq 10^{-5}$
Mpc$^{-3}$ for SCUBA sources. Figs. \ref{nsfr} and \ref{phivsz}
demonstrate that in `realistic' models such space densities are only
achieved for objects with much lower star formation rates of $~\sim
10^{2}$ $\textrm{M}_{\odot} \textrm{yr}^{-1}$.  The common solution of
making the star formation more `bursty' does not help us match these
high space densities unless star formation timescales are considerably
shorter than we have assumed. This argues in favour of lower star
formation rates of the SCUBA sources than those inferred by
\citet{2002MNRAS.331..817S} unless the star formation efficiency in
the SCUBA sources is extremely high.

\section{Spectral Energy Distributions}\label{SED}

\subsection{Grey-body models of the Far-Infrared SED} 

For a given illuminating radiation field, optical depth and age, the
spectral energy distribution of a galaxy in the far-infrared (FIR) and
sub-mm part of the spectrum is determined by the dust temperature and
grain size. This part of the SED is relatively featureless
(\citealt{1997A&A...326..950F}; \citealt{1998ApJ...509..103S};
\citealt*{1999A&A...350..381D}; \citealt*{2000MNRAS.313..734E}),
because in clouds that are optically thick to optical/UV radiation,
the dust emission can be qualitatively represented by a single large
grain model whose extinction efficiency changes with the inverse of
the grain size. The presence of grains of smaller size results in a
distortion of the grey-body SED of the single grain model. We write
the grey-body SED as $L_{\nu}(\nu)\propto \varepsilon_{\nu}
B_{\nu}(T,\nu)$, where $L_{\nu}$ is the spectral luminosity of the
galaxy, $B_{\nu}$ is the Planck function, and $\varepsilon_{\nu}
(\propto \nu^{\beta})$ is the dust emissivity function. In the simple,
single-component dust model, $\beta=1$. The  presence of smaller
grains results in $\beta>1$. Provided that the grain size distribution
is not changed through creation and destruction of different-sized
grains, $\beta$ depends only weakly on temperature. In this paper,
$\beta$ is taken as a free parameter of the model and the effective
temperature of the dust, T, is the second adjustable parameter.

\begin{figure}
\begin{center}
\includegraphics[width=1.0\columnwidth]{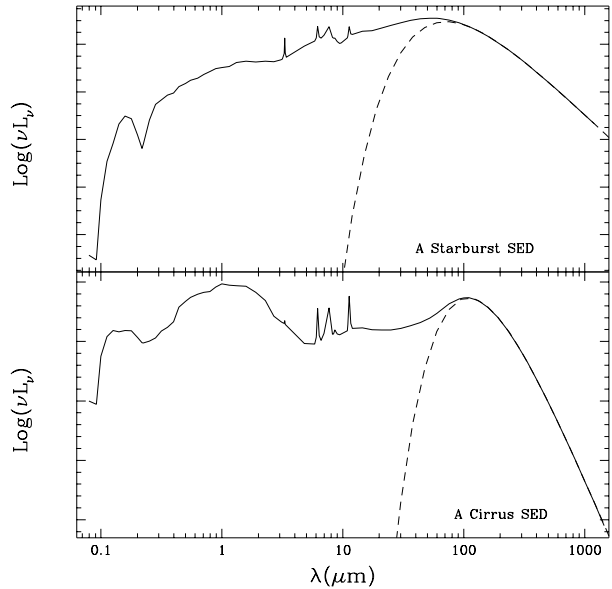}
\end{center}
\caption{Upper diagram: the SED of a starburst in a giant molecular
cloud calculated with the model of \citet{2000MNRAS.313..734E}. The
dashed line is a grey-body spectrum fitted to the section of the SED
with $\lambda>100$ $\mu$m. The parameters are $T_{s}= 34$ K, and
$\beta_{s}=1.9$. Lower diagram: the SED of a cirrus galaxy calculated
with the model of ERR. The parameters are $T_{s}= 22$ K, and
$\beta_{s}=1.9$.}
\label{sedfit}
\end{figure}

Fig. \ref{sedfit} shows grey-body fits to the SEDs of
\citet{2000MNRAS.313..734E} and ERR in the wavelength range
$\lambda\ge 100$ $\mu$m. The fit to the FIR emission is indeed very
good. We therefore take a simple grey-body parameterization for the
SED to predict the counts from our galaxy formation model. We have
compared counts calculated using SEDs from full radiative transfer
models with those calculated from grey-body fits. The results are in
agreement to within $\sim 1$ per cent.

The next step is to determine the normalization of the SED. We adopt
the common practice of relating the bolometric luminosity of a galaxy
to its instantaneous star formation rate through a linear
relationship, ignoring the cumulative effect of old stars, and thus
the star formation history of the source. We do this in order to keep
the number of dust parameters small. The form of the assumed initial
mass function (IMF), the evolution of the metallicity and the dust
mixture and the timescale of star formation all contribute to the
uncertainty in the proportionality constant of the model,
\begin{equation}
L_{bol}=\zeta\dot{M}.
\end{equation}

In the literature $\zeta \sim 10^{9}$ to $10^{10}$
$\textrm{L}_{\odot}\textrm{M}_{\odot}^{-1}\textrm{yr}$ for a range of
IMFs (see review by \citealt{astro-ph/0202228}). Here, we adopt a value
of $5\times 10^{9}$ and note that it is uncertain by a factor of two
in either direction. We further assume that the galaxies are optically
thick at visible and UV wavelengths and that most starlight is
absorbed and re-radiated by dust, i.e.  $L_{\rm fir}\sim L_{\rm bol}$.

It is customary to allow the high frequency limit of the slope of the
SED to be another parameter of a grey-body SED model. The effect of
this parameter in our calculation is very small since we are
interested in the flux far beyond the FIR peak. This parameter only
affects the normalization of the SED and its effect is a change in
$\zeta$ of $\lesssim 10$ per cent or a change of $\sim 1$ K in the
temperature. We shall therefore omit it in the interest of keeping the
number of parameters small.

\subsection{The sub-mm counts within hierarchical models with a hot
starburst-like SED}\label{hotcold}

In \S\ref{hierarchy}, we saw that, in hierarchical galaxy formation models,
the predicted space density of galaxies with star formation rates
$\gtrsim 100$ $\textrm{M}_{\odot}\textrm{yr}^{-1}$ is small. This is
the main reason for the difficulty encountered in fitting the sub-mm
source counts when assuming a hot SED. Higher dust temperatures move
the peak of the SED towards shorter wavelengths. Thus, for a given
bolometric luminosity of the source, the hotter the SED, the smaller
the proportion of the galaxy's output in the sub-mm. 

\begin{figure}
\begin{center}
\includegraphics[width=1.0\columnwidth]{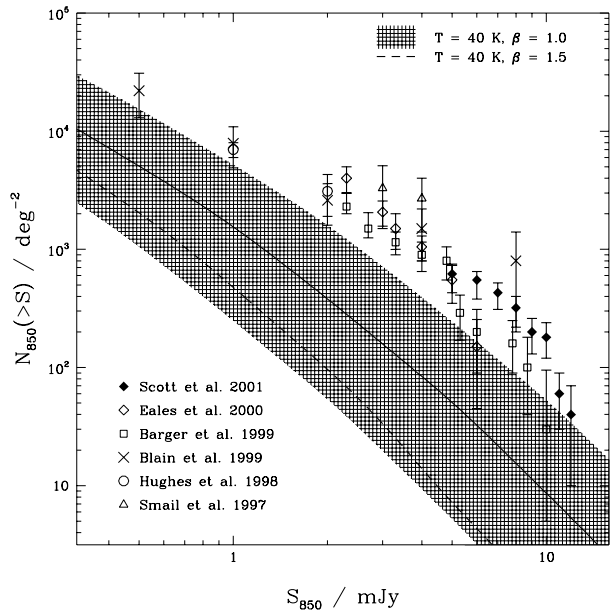}
\end{center}
\caption{Integral number counts at $\lambda=850$ $\mu$m. Hierarchical
models under-predict the counts at the bright end if a hot SED is
chosen for a starburst. $\zeta=5.0\times 10^{9}$
$\textrm{L}_{\odot}\textrm{M}_{\odot}^{-1}\textrm{yr}$, $t_{b}=3\times
10^{8}$ yr. In the shaded area around the solid line, $\zeta$ is
varied between $2.0\times 10^{9}$ and $10^{10}$
$\textrm{L}_{\odot}\textrm{M}_{\odot}^{-1}\textrm{yr}$}
\label{hot}
\end{figure}

To demonstrate this more clearly, the curves in Fig. \ref{hot} show
the sub-millimetre counts predicted for the semi-analytic models,
assuming that all galaxies have hot SEDs. Results are shown for
different values of the dust emissivity parameter, $\beta$. Even
though the temperature is at the lower end of the frequently assumed
range (40 -- 60 K), the predicted counts fall significantly short of
the observed values. Increasing either the temperature or the
emissivity index makes matters worse.

We saw in \S\ref{hierarchy} that the abundance of sources with
moderate star formation rates of the order $\dot{M} \sim$ 50 -- 100
$\textrm{M}_{\odot}\textrm{yr}^{-1}$ is sufficient to match the number
of objects detected in sub-mm surveys. For these moderate star
formation rates, it is unnecessary to assume a hot SED. In the next
section, we explore the effect of varying the dust temperature more
freely, while in \S\ref{LBG}, we will discuss the implications for the
relation of sub-mm sources to Lyman-break galaxies.

\subsection{Temperature as a free parameter in models with diffuse
extended dust}\label{freeT}

In the literature, a range of temperatures has been used when fitting
sub-mm counts, but most authors have refrained from allowing $T$ to
vary outside the range 40 -- 60 K, arguing that temperatures much
below this are unrealistic in a starburst galaxy, where star formation
takes place in optically thick giant molecular clouds (GMCs). In
standard radiative transfer models, the dust temperature of a GMC is
determined by its physical extent, which is in turn limited by the
time before it is dispersed by supernovae. The bolometric luminosity
of a galaxy then scales with the number of molecular clouds contained
within it.

\begin{figure}
\begin{center}
\includegraphics[width=1.0\columnwidth]{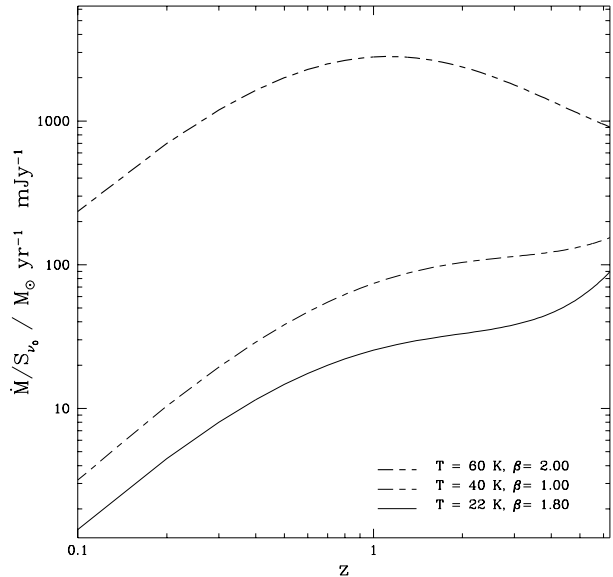}
\end{center}
\caption{The star formation rate of a source with a flux of 1 mJy at
850 $\mu$m for a range of hot and cold SEDs as a function of
redshift.}
\label{sfrs}
\end{figure}

For galaxies with moderate star formation rates, it is plausible  that
most of the IR emission is due to a different  geometrical
configuration of the star/dust distribution, more similar to that
responsible for the cirrus emission in spiral galaxies (ERR). If the
starlight is absorbed by diffuse extended dust and re-radiated in the
FIR, the intensity of the UV radiation heating the dust is not set by
the properties of GMCs and much lower dust temperatures are
expected. The peak in the FIR emission then moves much closer to 850
$\mu$m and the resulting  bolometric correction and the inferred star
formation rate decrease strongly.  Fig. \ref{sfrs} shows how the
inferred star formation rate for a source with a flux of 1 mJy at 850
$\mu$m varies as a function of redshift for a range of SEDs
corresponding to different dust temperatures.

\subsection{Further constraints on the dust models}\label{dust}

The 850-$\mu$m source counts are not able to constrain all the free
parameters needed to predict the sub-mm emission from the galaxies in
our model. We have thus chosen plausible values for two of our
parameters, $\zeta=5.0\times 10^{9}$
$\textrm{L}_{\odot}\textrm{M}_{\odot}^{-1}\textrm{yr}$ and
$t_{b}=3\times 10^{8}$ yr.

One way to break the degeneracy between star formation rate and
temperature is to consider the inferred size of the emitting region
(\citealt{2001ApJ...549..745R}). The predicted radius of the
emitting region depends on its bolometric luminosity and its surface
brightness as

\begin{equation}\label{r2}
4\pi r^{2}=\frac{L_{\rm fir}}{I_{\rm fir}}. 
\end{equation}
Surface brightness and temperature are linked because of the
equilibrium between absorbed UV/optical emission and emitted IR
emission, $I_{\rm fir} \propto I_{\rm heat}$. Higher UV/optical
surface brightness, $I_{\rm heat}$, will result in higher dust
temperatures which in turn will result in a higher IR surface
brightness. 

In order to normalize the relation between the intensity of the
radiation field and the emitted bolometric surface brightness, we have
to make assumptions about the relative distribution of dust and
stars. Observed sub-mm sources are generally very faint in the optical
and UV and are therefore heavily obscured at these wavelengths. We
will consider two limiting cases: a dust screen and an extended
distribution of stars and dust. For a dust screen, the situation is
basically the same as in the optically thin case. The dust layer
located within about one optical depth from the stars will be heated
to approximately the same equilibrium temperature as in the optically
thin case. The surface brightness of starlight needed to reach a
certain temperature is fixed and can be used to infer the bolometric
FIR surface brightness and hence the size of the IR-emitting region.

The dust properties (composition, grain size distribution) will fix
the dust emissivity index $\beta$. The predicted surface
brightness/size will increase/decrease with increasing star formation
rate. The relations between temperature, bolometric luminosity,
bolometric surface brightness and sub-mm flux can then be used to
predict the sizes of sources of given sub-mm flux in our models, or,
more generally, to infer a star formation rate from an observed sub-mm
flux and size. We will discuss this in more detail in \S\ref{siz}.

\section{Results}\label{fit}

\subsection{Fitting the 850-$\mu$m counts} 

\begin{figure}
\begin{center}
\includegraphics[width=1.0\columnwidth]{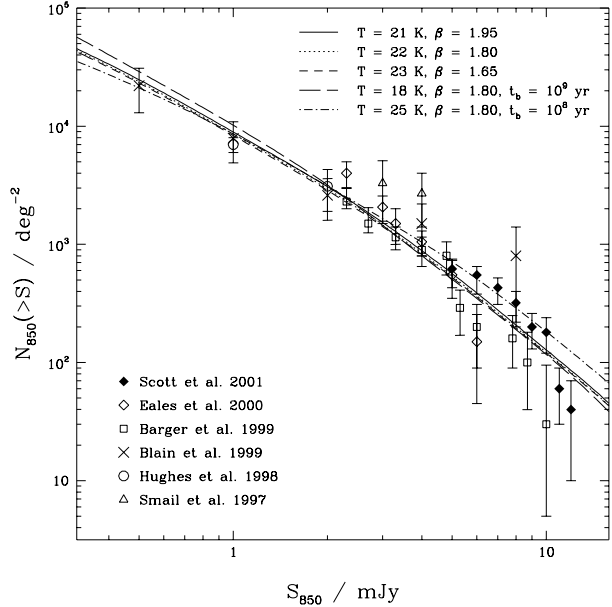}
\end{center}
\caption{Integral number counts at 850 $\mu$m. The sources of the data
are given in the legend. The error bars in each data set are
correlated. $\zeta=5.0\times 10^{9}$
$\textrm{L}_{\odot}\textrm{M}_{\odot}^{-1}\textrm{yr}$, $t_{b}=3\times
10^{8}$ yr unless otherwise stated.}
\label{counts}
\end{figure}

In this section, we use the star formation rates given by the
hierarchical galaxy formation model, described in
\S\ref{semianalytic}, to calculate the 850-$\mu$m counts. For
simplicity, we have assumed that the sub-mm sources are all
star-forming objects characterized by a single temperature SED. Of
course, in reality, sources of a range of temperatures, most likely
also belonging to different populations of sources (quiescently
star-forming galaxies, starbursts, AGN, etc.), will contribute to the
850-$\mu$m counts.

There is a certain amount of degeneracy between the temperature and
the emissivity index. As illustrated in Fig. \ref{counts}, the
temperature required to obtain a reasonable fit to the data changes
from 23 K to 21 K as we vary $\beta$ in the range 1.65 to 1.95. In
physical terms, the value of $\beta$ is set by the precise nature and
composition of the dust mixture, its optical depth, and how these vary
as a function of the age of the stars that are responsible for the
heating. Realistic radiative transfer models
(\citealt{1998ApJ...509..103S}; \citealt{2000MNRAS.313..734E}; ERR)
yield $\beta$ values in the range 1.5 -- 2.0.

The parameter $\zeta$ parameterizes the effect of the IMF on the
energy output by a given mass of stars that is formed. A change in
$\zeta$ moves the count distribution horizontally. A change of the
star formation time scale, $t_{b}$, alters both the luminosity of the
sources and the length of time they are `visible', so its effect on
the count distribution is more complex. The bright counts are
dominated by the most strongly star-forming objects. Because the star
formation rate function is very steep at the bright end (see
Fig. \ref{nsfr}), an increase in the luminosity of the sources tends
to over-compensate for the drop in density. The faint end of the star
formation rate function is shallow, so at fainter flux levels the
normalization of the counts is determined by the space density of the
sources.  The cross-over of the two regimes takes place at $\sim 0.1$
mJy. Over the range of observed source fluxes, a lower value of
$t_{b}$ will thus reduce the slope of the counts.

The effect of a scatter in the temperatures of the sub-mm emitting
regions is to reduce the slope of the flux distribution, as lower 
temperature sources will make a large contribution to the counts at 
the bright end \citep{astro-ph/0107290}.

\subsection{Sizes of the Source Galaxies}\label{siz}

Because of its large beam size, SCUBA is unable to constrain the sizes
of the detected sub-mm sources. The only information we have so far on
the sizes of these sources comes from observations with the Plateau de
Bure interferometer which appear to have marginally resolved the SCUBA
galaxy Lockman850.1 \citep{2001A&A...378...70L}. As we will discuss
below, the size of this galaxy fits in with the assumption of low
dust temperatures.

Let us define the SED of a source as $L_{\nu}=f_{\nu}L_{bol}$, where
$f_{\nu}$ is a normalized function. We can then use the flux redshift
relation to relate flux and luminosity,
\begin{equation}
\nu_{0}S_{\nu_{0}}=\frac{\nu L_{\nu}}{4\pi D_{L}^{2}}.
\end{equation}
Here, $D_{L}$ is the luminosity distance. Equation \ref{r2} then
gives an expression for the angular size of the object in terms of its
bolometric surface brightness $I_{\rm fir}$,
\begin{equation}\label{theta}
\Theta S_{\nu_{0}}^{-1/2}=(1+z)^{3/2}(f_{\nu}I_{\rm fir})^{-1/2}. 
\end{equation}

As discussed in \S\ref{dust}, for a given grain composition, the
balance between absorbed and emitted energy establishes a tight
relation between the intensity of the UV/optical radiation and the
temperature (distribution) of the dust. In the case of an extended,
optically thin dust distribution, this intensity will be given by the
average surface brightness of the starlight (as is the case for cirrus
emission from spiral galaxies). As discussed,  the resulting
FIR SED can be well fitted by a grey-body spectrum with an effective
temperature and $T^{4+\beta} \propto I_{\rm heat} $.

\begin{figure}
\begin{center}
\includegraphics[width=1.0\columnwidth]{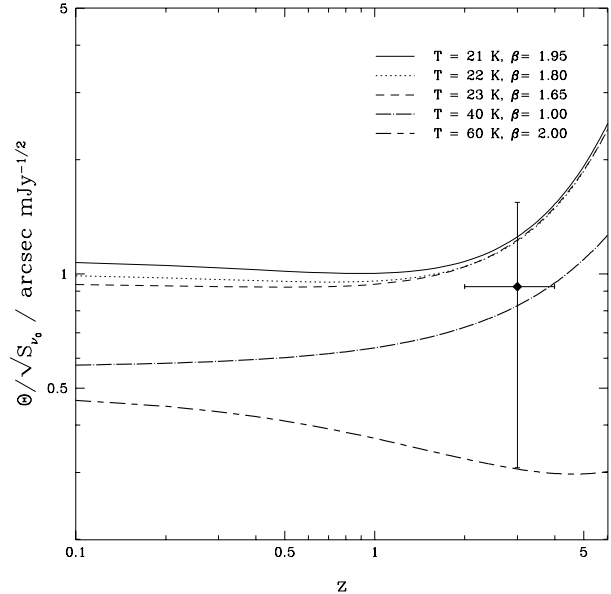}
\end{center}
\caption{The angular size of a source with a flux of 1 mJy at 850
$\mu$m for the fitting models and for the limiting cases in the
commonly assumed range of $40\textrm{ K}<T<60\textrm{ K}$ and
$1<\beta<2$. The data point is taken from \citet{2001A&A...378...70L}}
\label{angular}
\end{figure}

We have used the cirrus model of ERR to calculate the SED and the
relation between temperature and $I_{\rm heat}$. If the dust acts as a
screen $I_{\rm fir} \sim (1-e^{-A_{V}}) I_{\rm heat} \sim I_{\rm
heat}$, while for an extended distribution of dust and stars $I_{\rm
fir} \sim A_V I_{\rm heat}$. Fig. \ref{angular} shows the resulting
angular sizes as a function of redshift  for sources with
$S_{\nu_{0}}=1$ mJy, assuming an extended dust distribution with
$A_{V}=10$. Results are shown for a range of different dust
temperatures.  The value of $A_{V}$ we assume is slightly larger than
that adopted by ERR to fit the SEDs of non-starburst SCUBA
sources. However, in order to have optical data, the  sources they
consider are necessarily less obscured. Note the obvious scaling
$\propto S_{\nu_{0}}^{1/2}$. For a dust screen geometry, the predicted
sizes would be larger by a factor three.

The angular size of the sources is almost constant at low
redshifts. Once the k-correction no longer compensates for the
redshift factor in equation \ref{theta}, the angular sizes increase
with redshift. For low dust temperatures, the sizes are around
$1 (S_{850}/1\textrm{mJy})^{1/2}$ arcsec for redshifts less than $\sim
2$ and then increase by a factor $2$ in the redshift interval
$2<z<6$. We also show two models with temperature and emissivity
appropriate for hot starburst SEDs. At a given redshift, the angular
size of the source decreases approximately linearly with the assumed
temperature. The cool, extended models predict significantly larger
sizes than the hot SED models.  A galaxy with $(T,\beta)=(22\textrm{
K},1.8)$ would be two to three times as large as a hot source in the
redshift range $2<z<6$.

The weak redshift dependence of the angular sizes makes it possible to
use observed sizes to constrain the dust temperature in the screen
model and a combination of temperature and optical depth for the
extended model.

The cross in Fig. \ref{angular} shows the size of SCUBA galaxy
Lockman850.1 assuming that it is marginally resolved and scaled to an
850-$\mu$m flux of 1 mJy \citep{2001A&A...378...70L}. This observation
is in reasonably good agreement with our prediction for emission from
cold extended dust. Note, however, that LE850.6, resolved in the radio
seems to have a slightly smaller size \citep{astro-ph/0206432}. 

\begin{figure}
\begin{center}
\includegraphics[width=1.0\columnwidth]{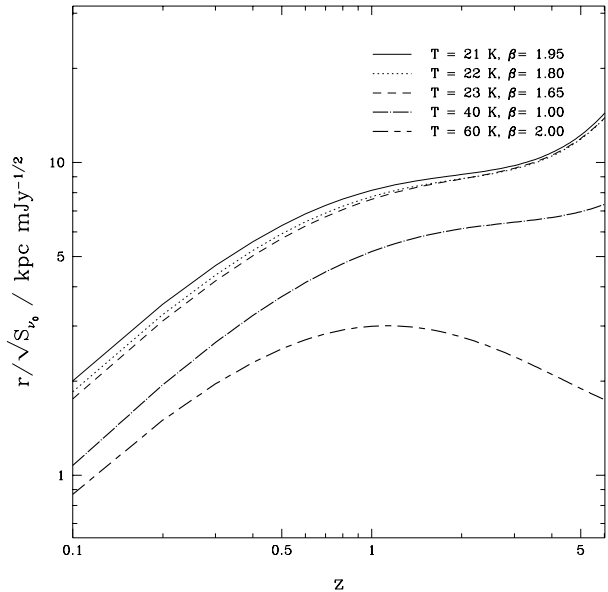}
\end{center}
\caption{The size of a source with a flux of 1 mJy at 850 $\mu$m for
the fitting models and for the limiting cases in the commonly assumed
range of $40\textrm{ K}<T<60\textrm{ K}$ and $1<\beta<2$.}
\label{sizes}
\end{figure}

The corresponding proper sizes of the sources depend on the
cosmology. With the same parameters as in Fig. \ref{counts}, our model
predicts sizes $\gtrsim 5$ kpc (Fig. \ref{sizes}).

Changes in $\zeta$ and $t_{b}$ require changes in the assumed dust
temperature and emissivity index in order to maintain a good fit to
the counts. An increase in $\zeta$ means that the dust temperatures
can be higher, since a smaller fraction of the stellar output is
needed in the sub-mm. Thus the sources can be more compact, with
larger surface brightnesses and smaller radii. Increasing the star
formation timescale has the opposite effect. The bolometric source
luminosities drop and cooler, more extended sources are required to
match the observed sub-mm fluxes.

\subsection{Redshift Distributions}

\begin{figure}
\begin{center}
\includegraphics[width=1.0\columnwidth]{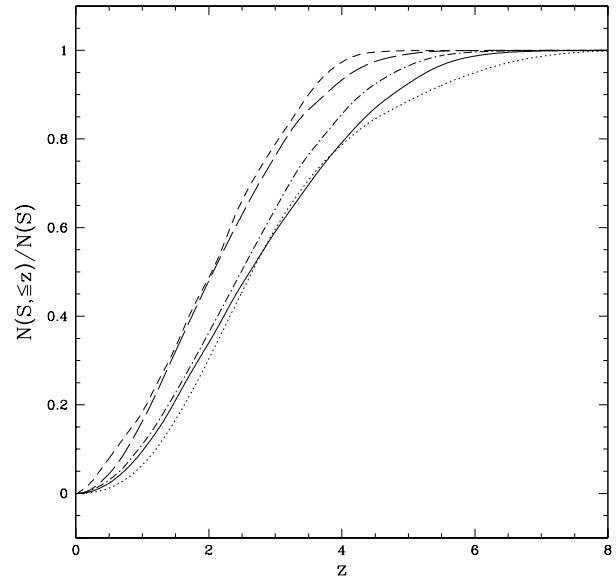}
\end{center}
\caption{Redshift distribution of the sources for the model with
$\zeta=5\times 10^{9}$
$\textrm{L}_{\odot}\textrm{M}_{\odot}^{-1}\textrm{yr}$ and
$t_{b}=10^{8}$ Gyr and $T=22$ K, at $S_{\nu_{0}}=1$ mJy (solid), $0.3$
mJy (dotted) and 10 mJy (short-dashed). The other two lines are both
at 1 mJy and $t_{b}=10^{9}$ Gyr, $\zeta=5\times 10^{9}$
$\textrm{L}_{\odot}\textrm{M}_{\odot}^{-1}\textrm{yr}$, $T=17$ K
(long-dashed) and $\zeta=8.9\times 10^{9}$
$\textrm{L}_{\odot}\textrm{M}_{\odot}^{-1}\textrm{yr}$, $T=21$ K
(dot-dashed). All have $\beta=1.8$}
\label{zdistribution}
\end{figure}

Fig. \ref{zdistribution} shows the redshift distribution of our
sources down to various flux limits for a range of different
parameters. The typical redshifts are in the range $2\lesssim
z\lesssim 4$. This is very similar to previous modelling which has
assumed hot starburst-like SEDs. 

The solid curve shows the redshift distribution down to a flux level
of 1 mJy with the parameter set $(T,\beta)=(22\textrm{ K},1.8)$. The
median redshift of the sources is 2.6. The mean of the redshift
distribution is 2.8 and its variance 1.4. The distribution has a high
redshift tail that extends to $z\approx 6$. The distribution of the
fainter sources is similar, but extends to larger redshifts (dotted)
while the bright sources peak at slightly lower redshifts
(short-dashed). The curves change little if we change the starburst
time scale and the assumed IMF (long-dashed and dot-dashed,
respectively)

Because the sub-mm counts are strongly biased towards high-redshift
sources, counts at shorter wavelengths place rather little constraint
on models for the sub-mm source population (see
e.g. \citealt{2000ApJ...542..710G}).

\section{Discussion}\label{discuss}

\subsection{The nature of sub-mm sources and the relation to
Lyman--break galaxies} \label{LBG}

As discussed in the previous section, low dust temperatures imply
a star/dust distribution that is spatially extended over $\gtrsim 5$
kpc. This is very different from typical violent starbursts at the
centre of galaxies which are spatially compact with typical sizes
$\lesssim 100$ pc. In our model, the sub-mm sources are more extended
than Lyman--break galaxies which have typical sizes of 0.6 arcsec
\citep{2001ApSSS.277..609C}.

The UV continuum slope of Lyman--break galaxies indicates that they
are typically optically thick in the optical/UV
\citep{2001ApJ...554..981P} and emit the bulk of their light in the
IR. Typical correction factors to the bolometric luminosity range from
3 to 15 depending on the assumed reddening curve
\citep{1998ApJ...492..428S}. Furthermore, the spectra of Lyman--break
galaxies exhibit many of the characteristics of nearby starbursts. It
was thus expected that they would dominate the sub-mm source
population and the lack of sub-mm detections of most Lyman--break
galaxies came as a  surprise (\citealt{1998Natur.394..241H};
\citealt{1998ApJ...507L..21S}; \citealt{1999ApJ...518L...5B};
\citealt{1999ApJ...518..641L}; \citealt{2000MNRAS.319..318C}).

The larger spatial extent and smaller surface brightness predicted by
our cold dust models may naturally explain the lack of overlap of the
sub-mm sources and Lyman--break galaxies. Optical/UV selected samples
of high-redshift galaxies will be strongly biased to high surface
brightness objects. These will have high dust temperatures and, at the
typical redshifts of these sources, the 850-$\mu$m flux will come from
far down the Raleigh-Jeans tail of the spectrum. Conversely, a larger
fraction of the total emitted light of low surface brightness objects
will be detected in the sub-mm. In this way, the UV to sub-mm flux
ratio can differ by a factor 30 -- 100 for objects with the same
optical depth and star formation rate (see Fig. \ref{sfrs}). A larger
optical depth in sub-mm detected sources would obviously further
increase this ratio (see \citealt*{2001MNRAS.327..895S} for a model
where the bright sources are more strongly obscured).

\subsection{Contribution from AGN}

Active Galactic Nuclei (AGN) make a contribution to the FIR and sub-mm
luminosities of Luminous IR Galaxies  by heating the dust in torus
enshrouding an AGN (e.g. \citealt{1995MNRAS.272..737R}). The debate on the
fraction of the infrared luminosity due to AGN has not yet been
settled, though recent surveys with {\it Chandra} seem to suggest that
the contribution of AGN to SCUBA sources is not significant
\citep{2001AJ....122.2156A}. Observations of galaxies in the local
universe suggest 70 -- 80 per cent of ULIRGs
($10^{12}\textrm{L}_{\odot}<L_{fir}<10^{13}\textrm{L}_{\odot}$) are
starburst-dominated \citep{1998ApJ...498..579G}. Only when the
luminosity reaches that of a Hyper-luminous IR Galaxy (HLIRG,
$L_{fir}>10^{13}\textrm{L}_{\odot}$), are AGN thought to dominate the
far-IR emission  \citep{astro-ph/0205422}. For almost all the sources
in our model $L_{IR}/\textrm{L}_{\odot}\lesssim 10^{12.3}$. The
average contribution from AGN to these galaxies is of order 1 per cent
\citep*{1999ApJ...522..113V}.

\subsection{Implications for source counts in the mid IR } 

The forthcoming Spectral and Photometric Image Receiver (SPIRE;
\citealt*{2001ESASP.460...37G}) on European Space Agency's {\it Herschel
Space Observatory} will image the sky at 250, 350, and 500 $\mu$m. The
high redshifts of our source population means that the first of these
channels already looks at emission from beyond the FIR peak of our
SEDs ($z\sim 1.2$). Only the brightest low-redshift SCUBA sources
would be detectable in a 250 $\mu$m SPIRE survey. At 350 $\mu$m, SCUBA
fluxes are almost unchanged for sources at the median redshift. The
500 $\mu$m channel will see most objects with $S_{850}\gtrsim 5$ mJy
over the whole redshift range. The point source sensitivity of the
channels is adequate to measure the slope of the SED, and hence the
temperature, of at least the brightest of the sources, directly. If
sources exhibit some scatter in temperature (as is likely), all the
above surveys will be biased toward the sources with higher
temperatures than we have assumed in this paper.

\section{Conclusions}\label{conculsions}

The sub-mm source counts predicted by hierarchical models of galaxies
formation have been studied for a wide range of SEDs. 

We confirm the results of previous studies that, with plausible
assumptions about star formation timescales, star formation efficiencies
and IMF, hierarchical models for galaxy formation under-predict the
number of bright sub-mm sources by a large factor if hot (40 -- 60 K)
starburst-like SEDs are assumed. 

The sub-mm source counts alone do not constrain the dust temperatures
and SEDs of the sources. Dust temperatures and inferred star formation
rates are highly degenerate. Lower dust temperatures require
substantially lower star formation rates to produce the same
850-$\mu$m flux.

For dust temperatures in the range 20 -- 25 K, the observed sub-mm
counts agree very well with those predicted by hierarchical galaxy
formation models. These low temperatures imply typical radii that
correspond to angular sizes of $1 (S_{850}/1\textrm{mJy})^{1/2}$
arcsec if we assume an extended distribution of dust and stars  with
$A_V=10$. This is a factor two to three larger than predicted for
starburst-like SEDs. Observational limits on the size of sub-mm
sources are scarce. Marginally resolved observations of the SCUBA
galaxy Lockman850.1 with the Plateau de Bure interferometer, and
follow up near infrared imaging, may indicate that the sub-mm sources
are as large as predicted by our cold extended dust models.  The
typical redshift of the sources should then lie in the range
$2\lesssim z\lesssim 4$, somewhat, but not significantly, smaller than
in models which assume a starburst-like SED. These low temperatures
substantially reduce the predicted mid-IR emission of the sub-mm
sources. If our models are correct, SCUBA-selected sources may
contribute less than expected to the counts in upcoming mid-IR
surveys, which will be dominated by sources with hotter dust.

The low temperatures required to reconcile the sub-mm counts with the
predictions of hierarchical galaxy formation models are not plausible
for compact star-bursting galaxies, but they are expected if the
emission comes from a more extended distribution of stars and dust. If
most sub-mm galaxies have indeed extended distributions of cold dust
with large sub-mm to UV/optical ratios this would naturally explain
the small overlap with the population of Lyman--break galaxies, which
is biased towards high surface brightness regions in actively
star-forming galaxies.

\section{Acknowledgements}
We thank Andreas Efstathiou for making his code for the modelling of
dust emission available, and the referee, J. E. G. Devriendt, for
his helpful comments and suggestions.

\bibliography{cold_5}

\end{document}